

\input harvmac
\def\rhob{{\rho\kern-0.465em \rho}}

\def\no{\noindent}
\def\o{\over}

\def\alp{\alpha}\def\del{\delta}

\def\ZZ{{\bf Z}}

\def\ontopss#1#2#3#4{\raise#4ex \hbox{#1}\mkern-#3mu {#2}}

\setbox\strutbox=\hbox{\vrule height12pt depth5pt width0pt}

\def\strut{\relax\ifmmode\copy\strutbox\else\unhcopy\strutbox\fi}

\nref\rrogone{L.J. Rogers, Proc. London Math. Soc. (series 1) 25 (1894) 318.}
\nref\rrogtwo{L.J. Rogers, Proc. London Math. Soc. (series 2) 16 (1917) 315.}
\nref\rschur{I. Schur, Berliner Sitzungberichte 23 (1917) 301.}
\nref\rrrh{L.J. Rogers, S. Ramanujan and G.H. Hardy, Proc. Camb.
 Phil. Soc. 19 (1919) 211.}
\nref\rhardy{G.H. Hardy, {\it Ramanujan} (Cambridge University
 Press, 1940).}
\nref\rkacone{V.G. Kac, Funct. Anal. Appl. 1 (1967) 328; 2
 (1968) 183; Math. USSR-Izvestia 2 (1968) 1271.}
\nref\rmoody{R.V. Moody, Bull. Am. Math. Soc. 73 (1967) 217; J.
 Algebra 10 (1968) 211; Can. J. Math. 21 (1969) 1432.}
\nref\rkactwo{V.G. Kac, {\it  Infinite dimensional Lie algebras}, third
 edition (Cambridge University Press, 1990).}
\nref\rgno{P. Goddard and D. Olive, Int. J. Mod. Phys. A1 (1986) 303.}
\nref\rschwinger{J. Schwinger, Phys. Rev. Letts. 3 (1959) 296;
 Phys. Rev. 127 (1962) 324; 130 (1963) 496, 800.}
\nref\rvir{M. Virasoro, Phys. Rev. D1 (1970) 2933.}
\nref\rff{B.L. Feigen and D.B. Fuchs, Funct. Anal. Appl. 17 (1983) 241.}
\nref\rrc{A. Rocha-Caridi, in: {\it Vertex Operators in Mathematics and
 Physics}, ed. J. Lepowsky et al (Springer, Berlin, 1985).}
\nref\rsug{H. Sugawara, Phys. Rev. 170 (1968) 1659.}
\nref\rkacthree{V.G. Kac, Funct. Anal. Appl. 8 (1974) 68.}
\nref\rdas{R. Dashen and Y. Frishman, Phys. Rev. D11 (1975) 2781.}
\nref\rkniz{V.G. Knizhnik and A.B. Zamolodchikov, Nucl. Phys. B247
(1984) 83.}
\nref\rgko{P. Goddard, A. Kent and D. Olive, Phys. Lett. 152B
 (1985) 88; Commun. Math. Phys. 103 (1986) 105.}
\nref\rfl{A.J. Feingold and J. Lepowsky, Adv. in Math. 29 (1978) 271.}
\nref\rkacpet{V.G. Kac and D.H. Peterson, Adv. in Math. 53
 (1984) 125.}
\nref\rkacwak{V.G. Kac and M. Wakimoto, Adv. in Math. 70 (1988) 156.}
\nref\rWalg{A.B. Zamolodchikov, Theor. and Math. Phys. 65 (1985) 1205;
 F.A. Bais, P. Bouwk\-ne\-ght, M. Surridge and K. Schoutens
 Nucl. Phys. B304 (1988) 348 and 371; V.A. Fateev and S.L. Lykyanov,
 Sov. Sci. Rev. A Phys. 15 (1990) 1.}
\nref\rabs{D. Altschuler, M. Bauer and H. Saleur, J. Phys. A23 (1990) L789.}
\nref\rjmone{M. Jimbo and T. Miwa, Adv. Stud. in Pure Math. 4 (1984)
 97.}
\nref\rlepmil{J. Lepowsky and S. Milne, Adv. in Math. 29
 (1978) 15.}
\nref\rlepwil{J. Lepowsky and R.L. Wilson, Proc. Nat. Acad. Sci. USA
 78 (1981) 7254; Adv. in Math. 45 (1982) 21.}
\nref\rleppri{J. Lepowsky and M. Primc, {\it Structure of the standard
 modules for the affine Lie algebra} $A_1^{(1)}$, Contemporary
 Mathematics, Vol. 46 (AMS, Providence, 1985).}
\nref\rlw{J. Lepowsky and R.L. Wilson, Invent. Math. 77 (1984) 199.}
\nref\rdl{C. Dong and J. Lepowsky, {\it Generalized vertex algebras
 and relative vertex operators}, to appear.}
\nref\rgor{B. Gordon, Amer. J. Math. 83 (1961) 393.}
\nref\rand{G.E. Andrews, Proc. Nat. Sci. USA 71 (1974) 4082.}
\nref\rfno{B.L. Feigin, T. Nakanishi and H. Ooguri, Int. J. Mod. Phys.
 A7, Suppl. 1A (1992) 217.}
\nref\rchri{P. Christe, Int. J. Mod. Phys. A6 (1991) 5271.}
\nref\rkrv{J. Kellendonk, M. R\"osgen and R. Varnahgen, Bonn preprint
 Bonn-HE-93-04, hep-th/9301086.}
\nref\rwittLG{E. Witten, Princeton preprint IASSNS-HEP-93/10,
 hep-th/9304026.}
\nref\rbethe{H. Bethe, Z. Phys. 71 (1931) 205.}
\nref\ronsager{L. Onsager, Phys. Rev. 65 (1944) 117.}
\nref\rbax{R.J. Baxter, Phys. Rev. Letts. 26 (1971) 832 and 834;
 Ann. Phys. 70 (1972) 323.}
\nref\rwannier{G.H. Wannier, Rev. Mod. Phys. 17 (1945),50; L. Onsager,
 in: {\it Critical phenomena in alloys, magnets and
 superconductors}, ed. R.E. Mills et al
 (McGraw-Hill, New York, 1971), pp. xi,3; H. Au-Yang and J.H.H. Perk,
 Adv. Stud. in Pure Math. 19 (1989) 57.}
\nref\ryang{C.N. Yang, Phys. Rev. Letts. 19 (1967) 1312;
 Phys. Rev. 168 (1968) 1920.}
\nref\rwmtb{T.T. Wu, B.M. McCoy, C.A. Tracy and E. Barouch, Phys. Rev.
 B13 (1976) 316; B.M. McCoy, C.A. Tracy and T.T. Wu, J. Math.
 Phys. 18 (1977) 1058.}
\nref\rbpz{A.A. Belavin, A.M. Polyakov and A.B. Zamolodchikov, J.
 Stat. Phys. 34 (1984) 763, and Nucl. Phys. B241 (1984) 333.}
\nref\rcardy{J.L. Cardy, Nucl. Phys. B270 [FS16] (1986) 186.}
\nref\rciz{A. Cappelli, C. Itzykson and J.-B. Zuber, Nucl. Phys. B280
[FS18] (1987) 445; D. Gepner, Nucl. Phys. B287 (1987) 111.}
\nref\rjw{P. Jordan and E. Wigner, Z. Phys. 47 (1928) 631.}
\nref\rkauf{B. Kaufman, Phys. Rev. 76 (1949) 1232.}
\nref\rcol{S. Coleman, Phys. Rev. D11 (1975) 2088.}
\nref\rman{S. Mandelstam, Phys. Rev. D11 (1975) 3026.}
\nref\rmel{T.R. Klassen and E. Melzer, Int. J. Mod. Phys. A (in press).}
\nref\rluther{A. Luther, Phys. Rev. B14 (1976) 2153.}
\nref\rluscher{M. L\"uscher, Nucl. Phys. B117 (1976) 475.}
\nref\rfrishman{Y. Frishman and J. Sonnenschein, Weizmann Institute
 preprint WIS-92/54/june-PH, hep-th/9207017.}
\nref\rfrenkel{I.B. Frenkel, J. Funct. Anal. 44 (1981) 259.}
\nref\rwit{E. Witten, Commun. Math. Phys. 92 (1984) 455.}
\nref\rkm{R. Kedem and B.M. McCoy, J. Stat. Phys. (in press).}
\nref\rdkmm{S. Dasmahapatra, R. Kedem, B.M. McCoy and E. Melzer,
 Stony Brook preprint ITP-93-17, hep-th/9304150.}
\nref\rkkmmone{R. Kedem, T.R. Klassen, B.M. McCoy and E. Melzer,
 Phys. Lett. B (in press), hep-th/9211102.}
\nref\rkkmmtwo{R. Kedem, T.R. Klassen, B.M. McCoy and E. Melzer,
 Phys. Lett. B (in press), hep-th/9301046.}
\nref\rdkkmm{S. Dasmahapatra, R. Kedem, T.R. Klassen, B.M. McCoy and
 E. Melzer, to appear in the proceedings of the conference
 ``Yang-Baxter equations in Paris'' (World Scientific),
  hep-th/9303013.}
\nref\rstanley{R.P. Stanley, {\it Ordered Structures and Partitions},
 Mem. Amer. Math. Soc. 119 (1972).}
\nref\rslater{L. Slater, Proc. Lond. Math. Soc. (series 2) 54 (1951-52)
 147.}
\nref\rwz{J. Wess and B. Zumino, Phys. Lett. B37 (1971) 95.}
\nref\raffleck{I. Affleck, Phys. Rev. Lett. 55 (1985) 1355.}
\nref\rft{L.D. Faddeev and L.A. Takhtajan, Phys. Lett. 85A (1981)  375.}
\nref\randrewstwo{G.E. Andrews, {\it The Theory of Partitions}
 (Addison-Wesley, Reading Mass., 1976).}
\nref\rkirits{E.B. Kiritsis, Phys. Lett.B217 (1989) 427.}
\nref\rgin{P. Ginsparg, Nucl. Phys. B295 [FS21](1988) 153.}
\nref\ryy{C.N. Yang and C.P. Yang, Phys. Rev. 150 (1966) 321 and 327;
  151 (1966) 258.}
\nref\rjkm{J.D. Johnson, S. Krinsky and B.M. McCoy, Phys. Rev. A8
 (1973) 2526.}
\nref\rter{M. Terhoeven, Bonn preprint Bonn-HE-92-36, hep-th/9211120.}
\nref\rkns{A. Kuniba, T. Nakanishi, and J. Suzuki, Harvard preprint
 HUPT-92/A069.}
\nref\rZF{A.B. Zamolodchikov and V.A. Fateev, Sov. Phys. JETP 62 (1985)
 215.}
\nref\rgep{D. Gepner, Nucl. Phys. B296 (1988) 757.}
\nref\rrisz{B. Richmond and G. Szekeres, J. Austral. Soc. (Series A)
 31 (1981) 362.}
\nref\rnrt{W. Nahm, A. Recknagel and M. Terhoeven, Bonn preprint,
 hep-th/9211034.}
\nref\rlew{L. Lewin, {\it Dilogarithms and associated functions}
 (MacDonald, London, 1958).}
\nref\rzam{A.B. Zamolodchikov,
 Adv. Stud. in Pure Math. 19 (1989) 641.}
\nref\rvafa{S. Cecotti, P. Fendley, K. Intriligator, and C. Vafa, Nucl.
Phys. B386 (1992) 405; P. Fendley and H. Saleur, Nucl. Phys. B388
(1992) 609; S. Cecotti and C. Vafa, Harvard preprints HUTP-92/A044 and
92/A064.}
\nref\rbaxtwo{R.J. Baxter, J. Stat. Phys. 26 (1981) 427.}
\nref\rabf{G.E. Andrews, R.J. Baxter and P.J. Forrester, J. Stat.
Phys. 35 (1984) 193.}
\nref\rfb{P.J. Forrester and R.J. Baxter, J. Stat. Phys. 38 (1985) 435.}
\nref\rhaldane{F.D.M. Haldane, Phys. Rev. Letts. 67 (1991) 937.}
\nref\rzou{Z. Zou and P.W. Anderson, Phys. Rev. B37 (1988) 627; S.A.
 Kivelson, D.S. Rokhsar and J.P. Sethna, Phys. Rev. B35 (1987) 8865.}
\nref\rlaughlin{R.B. Laughlin, Phys. Rev. Letts. 50 (1983) 1395.}
\nref\rgross{D.J. Gross, Princeton preprint PUPT-1355, hep-th/9212148.}

\Title{\vbox{\baselineskip12pt\hbox{ITP-SB-93-19}
\hbox{hep-th/9304056}}}
{\vbox{\centerline{The sums of Rogers, Schur and Ramanujan}
\vskip9pt
\centerline{ and the Bose-Fermi correspondence in}
\vskip9pt
\centerline{ $1+1$-dimensional quantum field theory}}}

\centerline{ Rinat~Kedem,
{}~Barry~M.~McCoy,
{}~and ~Ezer~Melzer~\foot{rinat ~or ~mccoy ~or melzer~@max.physics.sunysb.edu}}

\bigskip\centerline{\it Institute for Theoretical Physics}
\centerline{\it State University of New York}
\centerline{\it  Stony Brook,  NY 11794-3840}

\vskip 13mm

\centerline{\bf Abstract}
\vskip 4mm
We discuss the relation of the two
types of
sums in the Rogers-Schur-Ramanujan
identities with the Bose-Fermi correspondence of massless quantum
field theory in $1+1$ dimensions.
One type, which generalizes to sums which appear in the Weyl-Kac character
formula for representations of affine Lie algebras and in expressions
for their branching functions, is related to
bosonic descriptions of the spectrum of the field theory
(associated with the Feigin-Fuchs construction in conformal field theory).
Fermionic descriptions of the same spectrum are obtained via
generalizations of the other type of sums.
We here summarize recent results for such fermionic sum representations
of characters and branching functions.

\vskip 3mm

\Date{\hfill 4/93}
\vfill\eject

\newsec{Introduction}

The most important and difficult course taught in any university is
philosophy. It is of overriding importance because everyone has a
deeply rooted philosophy that rules
their
actions with an iron hand. It
is the most difficult of all courses to learn because no two people
are in complete agreement as to their philosophic principles. The
inevitable response to a question in philosophy is that it is either
trivially obvious or absolutely absurd. Unhappily, there is no agreement
about what is obvious and what is absurd. The consequence is that
almost nobody studies philosophy.

The second most difficult subject is physics. The difficulty is that
the study of physics requires students to hold two competing
philosophies in
their
minds at the same time and to form a synthesis. The
two competing philosophies are empiricism on the one hand, as embodied in
experiment and measurement, and rationalism or abstraction on the other
hand, as embodied in mathematics and computation. Physics is neither
the one nor the other but the Hegelian synthesis of both. More
students take elementary courses in physics than study Aristotle,
Acquinas, Kant and Hegel, but most do poorly and get bad grades.

\medskip
In the past (say) 30 years great progress has been made in theoretical
(or mathematical) physics. Yet because physics is a synthesis, the true
understsanding of the accomplishment is best made not by presenting
one set of developements but rather by describing two
parallel sets of developements, one loosely called mathematical and
the other loosely called physical. The mathematical side
of the developments we concentrate upon
embodies
Rogers-Ramanujan identities, modular forms,
infinite-dimensional algebras (such as affine Kac-Moody and the Virasoro
algebras)
and their representation theory. The physical side
involves statistical mechanics,
quantum spin chains,
quantum field theories (both conformal and massive), bosons and fermions.

Our ultimate goal here is to present the status of some of our recent
results on
fermionic sum representations for conformal field theory characters.
 This is done in section 4. However, we
also wish to elucidate the position our results have within the larger
tapestry of work of the last century in physics and mathematics. To
that end we will present
in sections 2 and 3, respectively,
the elements of the mathematical and the
physical sides out of which the synthesis is born.
The elements have
the names given in the previous paragraph. There is no accepted name for
the synthesis that puts both the competing elements on an equal
footing. We trust that this lack of a name will not be an impediment
to the philosophic reader.

\bigskip
\newsec{The mathematical ingredients}

In 1894 Rogers~\rrogone~proved the following set of identities
\eqn\szero{S_0=\sum_{n=0}^{\infty}{q^{n^2}\over
(q)_n}=\prod_{n=1}^{\infty}{1\o (1-q^{5n-1})(1-q^{5n-4})}
={1\over (q)_{\infty}}\sum_{n=-\infty}^{\infty}(q^{n(10n+1)}-q^{(5n+2)(2n+1)})}
\eqn\sone{S_1=\sum_{n=0}^{\infty}{q^{n(n+1)}\over
(q)_n}=\prod_{n=1}^{\infty}{1\o (1-q^{5n-2})(1-q^{5n-3})}
={1\over (q)_{\infty}}\sum_{n=-\infty}^{\infty}(q^{n(10n+3)}-q^{(5n+1)(2n+1)})}
where
\eqn\qdef{(q)_0=1~~,~~~~~~~  (q)_{n}=\prod_{k=1}^n(1-q^k)~~~~~~{\rm for}~~
  n=1,2,3,\ldots.}
A second proof by Rogers~\rrogtwo~and two independent proofs by
Schur~\rschur~were given in 1917. Hardy tells us~\rrrh \rhardy~that
the equality of the
left-hand sums with the products
was independently
conjectured by Ramanujan in 1913,
with a proof due to him published in~\rrrh, and
these equalities have subsequently
come to be known as the Rogers-Ramanujan identities.
There seems to be
no commonly accepted term which refers to all three expressions
in these identities of Rogers, Schur, and Ramanujan on the same footing.

The
products and the right-hand sums
in~\szero-\sone~may be directly
expressed in terms of theta functions~\rrogone \rrrh~and consequently
it is readily seen that if
one sets
\eqn\czero{c_0(q)=q^{-1/60}S_0(q)~~,~~~~~\quad c_1(q)=q^{11/60}S_1(q)}
and defines  $\tau$ by $q=e^{2\pi i \tau}$,
the   following  linear
transformation law is obtained:
\eqn\mod{\left(\matrix{c_0(-1/\tau)\cr c_1(-1/\tau)\cr}\right) =
{2\o\sqrt5}\left(\matrix{\sin{2\pi\o5} & \sin{\pi\o5}\cr
\sin{\pi\o5} & -\sin{2\pi\o5}\cr}\right)
\left(\matrix{c_0(\tau)\cr c_1(\tau)}\right)~.}
This enables one
to show that
$c_0(\tau)$ and $c_1(\tau)$ form a two-dimensional
representation of
the modular group. This group  has two generators
\eqn\gen{T:~\tau\rightarrow \tau+1~,~~~~~\quad S:~\tau\rightarrow -1/\tau~,}
which satisfy the relations
\eqn\relations{S^2=(ST)^3=1~.}

The second mathematical ingredient we need is the infinite-dimensional
generalization of Lie algebras introduced by Kac~\rkacone~and
Moody~\rmoody~in 1967. Our purpose here is not to review this
theory, which is presented in detail in~\rkactwo, but merely to recall
a few definitions.
In particular, given a simple Lie algebra $G$ of rank $r$
and dimension $d$
with
structure constants $f^{abc}$,
the untwisted affine  Lie algebra $G^{(1)}$
(defined in terms of a generalized Cartan matrix~\rkacone\rmoody)
is realized
by  the commutation relations
\eqn\km{[J_m^a,J_n^b]=\sum_{c=1}^d i f^{abc}J_{m+n}^c
  +k m \delta ^{ab}\delta _{m,-n}
  ~~~~~~(m,n\in \ZZ,~~a,b=1,\ldots,d),}
where we use the basis and
normalization conventions of~\rgno.
Here $k$ is a central element, {\it i.e.}~it commutes with every element of
the algebra, and  takes on a constant value in any given
irreducible representation of $G^{(1)}$.
The value of $2k/\psi^2$, where $\psi$ is the highest root of $G$
(which will be normalized to $\psi^2=2$ below), is
then called the level of the representation and is a positive integer
in the representations considered here.

It is worth noticing that the synthesis of the physical and the
mathematical is already inherent
in~\km.
This realization of the affine Kac-Moody
algebras, which was
derived from the definitions of~\rkacone\rmoody~in the late
1970s, had been
found earlier by Schwinger~\rschwinger~in his analysis of
relativistic invariance of gauge theories in 3+1
dimensions.
Consequently the central
element in \km, which plays a crucial role in
physics applications as well as in representation theory,
is at times~\rgno~referred
to as a Schwinger term.

Shortly after the
construction
of Kac-Moody algebras the Virasoro
algebra was introduced in 1970~\rvir.
This algebra is defined by the commutation relations
\eqn\vir{[L_m,L_n]=(m-n)L_{m+n}+{c\over 12}m(m^2-1)\delta_{m+n,0}
   ~~~~~~~~(m,n\in \ZZ),}
where the normalization of the $L_n$ is chosen such that
\eqn\lzero{[L_{\pm 1},L_0]={\pm}L_{\pm 1}~,~~~~~\quad[L_1,L_{-1}]=2L_0~~,}
and $c$ is  a central element
whose constant value in an irreducible representation is
called the central charge.

Of great importance are the   Virasoro   characters
\eqn\char{\chi_l(q)~=~ q^{-c/24}~{\rm Tr}~      q^{L_0}~~,}
where the trace is over an irreducible highest-weight representation
${\cal V}_l(c,\Delta_l)$ of the Virasoro algebra,
and the factor $q^{-c/24}$ is inserted to guarantee
linear behavior under the modular tramsformations \gen.
Such a representation ${\cal V}_l$,
and hence its character, is characterized by the central charge $c$ and
the highest
weight $\Delta_l$ which is the $L_0$-eigenvalue of the highest-weight
vector of ${\cal V}_l$.

Of interest to us here are several cases of these characters, as well
as characters of representations of various algebras which contain the Virasoro
algebra as a subalgebra (such as superconformal algebras, ${\cal W}$-algebras,
parafermionic algebras, and the already mentioned affine Lie algebras).
The first case concerns the irreducible representations of the Virasoro
algebra at central charge
\eqn\ccharge{c=1-{6(p-p')^2\over pp'}}
(where $p$ and $p'$ are coprime positive integers) and highest weights
\eqn\del{\Delta_{~r,s}^{(p,p')}={(rp'-sp)^2-(p-p')^2\over 4pp'}\quad
{}~~~~~(r=1,\ldots,p-1;\quad s=1,\ldots , p'-1).}
Based on the work of Feigin and Fuchs~\rff, Rocha-Caridi~\rrc~obtained the
following expressions for the corresponding characters:
\eqn\roc{
\widehat{\chi}_{~r,s}^{(p,p')} \equiv
q^{{c/24}-\Delta_{~r,s}^{(p,p')}}\chi _{~r,s}^{(p,p')}={1\over
(q)_{\infty}}\sum_{k=-\infty}^{\infty}(q^{k(kpp'+rp'-sp)}-q^{(kp'+s)(kp+r)})~.}
As required from $\Delta_{~r,s}^{(p,p')}=\Delta_{p-r,p'-s}^{(p,p')}$,
these characters have the symmetry
\eqn\sym{\chi_{~r,s}^{(p,p')}=\chi_{p-r,p'-s}^{(p,p')}~.}
We note in particular that if $(p,p')=(2,5)$, then the two
independent sums on the right-hand side of~\roc,
namely with $(r,s)$ set to (1,2) and (1,1),
are identical with
the two sums on the right-hand side of the Rogers-Schur-Ramanujan
identities~\szero\ and \sone, respectively.

The second case of interest here is that of the affine Lie algebras $G^{(1)}$.
Now $L_0$, entering the definition of
the characters~\char\ of level $k$ representations of $G^{(1)}$,
is quadratic in the generators $J_n^a$ of $G^{(1)}$. In fact, all the Virasoro
generators $L_n$ can be obtained from the $J_n^a$ via~\rgno \rdas-\rgko~the
 construction used by Sugawara~\rsug~in the analysis of Schwinger
terms~\rschwinger~in non-abelian gauge theory:
\eqn\kacvir{L_n~=~{1\over 2(k+g)}~\sum_{a=1}^r~ \sum_{m=-\infty}^{\infty}
 :J_{m+n}^a J_{-m}^a:~,}
where $g$ is the dual Coxeter number~\rgno~of $G$,
and the normal ordered
product :$J_m^a J_n^b$: equals $J_m^a J_n^b$ ~if $m\leq n$ and
$J_n^b J_m^a$ otherwise.   The corresponding Virasoro central charge is
\eqn\cGk{ c= {k ~{\rm dim}(G) \o k+g}~~.}
For a general algebra $G^{(1)}$ and arbitrary integer level $k$,
the characters \char,
where now the trace is taken over irreducible highest-weight
representations of $(G^{(1)})_k$,
 are given by the Weyl-Kac formula~\rkactwo \rkacthree.
In the case of $G=A_1\equiv su(2)$ and $k=1$ there are
two characters,
which can be written in the particularly simple form (cf.~\rfl)
\eqn\sucharacters{q^{1/24-l^2/4} \chi_l={1\over
(q)_{\infty}}\sum_{n=-\infty}^{\infty} q^{n(n +l)}~~~~~~~~\quad
 (l=0,1).}

The most general case we need are the characters of the algebras which arise
in coset constructions, introduced by Goddard, Kent and Olive~\rgko.
The characters are then branching functions of
affine Lie algebras~\rkacpet \rkacwak.
A wide class of cosets is given by
${(G^{(1)})_k\times(G^{(1)})_l\over  (G^{(1)})_{k+l}}$; for $l=1$
the corresponding branching functions are characters of the
${\cal W}G$-algebra~\rWalg~which reduces to the Virasoro algebra when $G=A_1$.
As a particular example,
the branching functions
for ${(A_{N-1}^{(1)})_1\times(A_{N-1}^{(1)})_1\over
(A_{N-1}^{(1)})_2}$~(which is equivalent by level-rank
duality~\rabs~to the
coset ${(A_1^{(1)})_N\over U(1)}$) are~\rkacpet \rjmone
\eqn\branch{\eqalign{
 q^{c/24-h_m^l} b_m^l =   {1\over (q)_{\infty}^2}&
 \Bigg[\Bigg(\sum_{s\geq0}\sum_{n\geq0}-\sum_{s<0}\sum_{n<0}\Bigg)
 (-1)^s q^{{s(s+1)\o 2}+(l+1)n+{(l+m)s\o 2}+(N+2)(n+s)n}\cr
 +& ~\Bigg( \sum_{s>0}\sum_{n\geq0}-\sum_{s\leq0}\sum_{n<0}\Bigg)
 (-1)^s q^{{s(s+1)\o 2}+(l+1)n+{(l-m)s\o 2}+(N+2)(n+s)n}\Bigg]~,\cr}}
where
\eqn\paracharge{c={2(N-1)\over N+2}~~,~~~~~~~~~~
   h_m^{l}={l(l+2)\over 4(N+2)}-{m^2\over 4N}~~.}
Here $l=0,1,\ldots,N-1$, ~$l-m$ is even, and the formulas are valid for
$|m| \leq l$ while for $|m|>l$ one uses the symmetries
\eqn\sym{b_m^l=b_{-m}^l=b_{m+2N}^l=b_{N-m}^{N-l}~.}

We note that
the right-hand sides
of~\roc,~\sucharacters, and~\branch\
share the feature with the sums on the right of~\szero-\sone~that
the denominator is a power
of $(q)_{\infty}$ and the numerator is a power series in $q$
(with integer coefficients).
Divided by the explicit power of $q$ on the left-hand sides,
they also share the property with~\czero~that they
can be seen to
form representations of
the modular group~\gen.
These features hold for
the general case of the Weyl-Kac character formula and the
branching functions as obtained from it.

\medskip
The above
discussion shows
that the right-hand side of
the Rogers-Schur-Ramanujan identities~\szero-\sone\
 has a vast generalization in
terms of characters of
representations of
infinite-dimensional
algebras. It is thus natural to ask whether the
remaining parts of these identities can also be generalized,
thus yielding different expressions for such characters.

The first step in this direction was taken by Lepowsky and
Milne~\rlepmil~in 1978 when they showed that for $A_1^{(1)}$ and
$A_2^{(2)}$ the Weyl-Kac formula, when suitably specialized,
admits
a product form. In 1981 Lepowsky and
Wilson~\rlepwil~found a way to obtain the sums on the left-hand side
of~\szero-\sone~using a construction which they called $Z$-algebras,
and thus provided a Lie-algebraic proof of
the Rogers-Ramanujan identities.

A major generalization of these results is due to Lepowsky
and Primc~\rleppri.
Extending the work of~\rlepwil\rlw~on $Z$-algebras (for a recent
review see~\rdl), they
found in 1985
that the branching
functions~\branch\ can be written as
\eqn\lpsum{
 q^{c/24}~q^{-{l(N-l)\over 2N(N+2)}}b_{2Q-l}^l~=
 \sum_{\scriptstyle m_1,\ldots ,m_{N-1}=0\atop \scriptstyle {\rm restrictions}}
 ^\infty {q^{{\bf m}C_{N-1}^{-1}{\bf m}^t-{\bf A}_l\cdot{\bf m}}\over
 (q)_{m_1} \ldots (q)_{m_{N-1}}}~~,}
where
$Q$ is an integer (mod $N$),
${\bf m}=(m_1,\ldots,m_{N-1})$ is subject to the restriction
\eqn\rest{\sum_{a=1}^{N-1}a m_a\equiv Q~({\rm mod}~N),}
$C_{N-1}$ is the Cartan matrix of the Lie algebra $A_{N-1}$
in the basis where
\eqn\qfAn{  {\bf m} C_{N-1}^{-1} {\bf m}^t ~=~
  {1\over N} \left( \sum_{a=1}^{N-1} a(N-a)m_a^2
 + 2\sum_{1\leq a < b \leq N-1}
     a(N-b) m_a m_b \right)~,}
and
${\bf A}_0$=0 while for $l=1,\ldots,N-1$
\eqn\lin{{\bf A}_l\cdot{\bf m}= -({\bf m}C_{N-1}^{-1})_l =
 -\left( {N-l\over N}\sum_{a=1}^{l} a m_a
 +{l\over N}\sum_{a=l+1}^{N-1}(N-a)m_a\right)~.}
This representation is of the form of a $q$-series which generalizes the
left-hand sums of~\szero~and~\sone\
to multiple sums
such as appear in the Gordon-Andrews identities~\rgor \rand.

Moreover, the sums in the Gordon-Andrews identities themselves have
also been found~\rfno~to be the Virasoro characters
\roc\ with $(p,p')=(2,2n+3)$.
(We note that the analysis in~\rfno~of the corresponding representations
of the Virasoro algebra leads directly to the product rather than the
sum sides of the Gordon-Andrews identities.)
In particular,
$\widehat{\chi}_{1,n+1}^{(2,2n+3)}$
is given by the
right-hand side of
\lpsum\ with no
restrictions on the sum, ${\bf A}_l=0$ and $C_{N-1}^{-1}$ replaced by
$(C'_n)^{-1}$, where $C'_n$
differs from $C_n$ only in one
entry which is $(C'_n)_{nn}=1$.
All the other characters are obtained~\rfno~by adding
suitable linear terms to the quadratic form in \lpsum, leading to the
full set of sums appearing in the Gordon-Andrews identities~\rgor\rand.
When $n$=1 one has $(p,p')$=$(2,5)$ and as noted above the
Virasoro characters reduce to the
original sums on the left-hand side of the Rogers-Ramanujan
identities~\szero-\sone.

However, until quite recently these were the only results known. The major
purpose of this note is to summarize the recent progress in finding
generalizations of the left-hand side of \szero-\sone~for the
Virasoro
characters~\roc,  the Weyl-Kac characters, and characters of the
coset models discussed above.
(Product formulas for characters have been recently discussed
in~\rchri -\rwittLG.)

\bigskip
\newsec{The physical ingredients}

There are at least three physical starting points which will be used
to form the synthesis with the mathematics of the previous section:
two-dimensional classical statistical mechanics,
one-dimensional quantum
spin chains,
 and conformal field theory.
The latter two
lead to the concept of
boson and fermion and to a relation between them that exists in 1+1
dimensions.

Consider first an $M$-body quantum
spin chain with periodic boundary conditions.
In the study of the spectra of $M$-body hamiltonians with local
interactions and translational invariance, the eigenstates which lie a
finite energy above the ground state energy as $M\rightarrow \infty$
have the quasi-particle form for the energy
\eqn\equasi{E_i-E_{GS}=\sum_{\alpha}~\sum_{j=1,{\rm rules}}^{m_\alpha}
e_{\alpha}(P_j^{\alpha})}
and for the momentum
\eqn\pquasi{P_i\equiv \sum_{\alpha}~\sum_{j=1,{\rm rules}}^{m_{\alpha}}
  P_j^{\alpha}~~({\rm mod}~2\pi),}
where $e_{\alpha}(P_j^{\alpha})$ is called the single-particle
excitation energy of type $\alpha$, and $m_{\alpha}$ are the numbers of
such excitations
in the eigenstate under consideration, which is labeled by $i$.
The sum over $P_j^{\alpha}$ is subject to certain
rules. If one of these rules is
\eqn\fermi{P_j^{\alpha}\neq P_k^{\alpha}~~~~\quad{\rm for}\quad j\neq k
  \quad{\rm and~all}\quad \alpha,}
the spectrum is called fermionic. If
there is no such exclusion rule
and an arbitrary number of
coinciding
$P_j^\alpha$ is allowed,
then
the spectrum is called bosonic.

The calculation of single-particle energies is extensively considered
in condensed matter physics. When considered on a lattice
they are
often periodic functions defined in an appropriate Brillouin zone.
 By definition
$e_{\alpha}(P)$ can never be negative. If all the $e_{\alpha}(P)$ are
positive the system is said to have a mass gap. If some
$e_{\alpha}(P)$ vanishes at some momentum
(say 0)
the system is said to be
massless, and for $P\sim 0$ a typical behavior is
\eqn\lindisp{e_{\alpha}(P)=v_\alpha |P|}
where $v_\alpha$ is variously referred to as the fermi velocity, spin-wave
velocity, speed of sound or speed of light.

In the statistical mechanics of many-body systems the most fundamental
quantity is the partition function which is defined as
\eqn\part{Z=\Tr\ e^{-{H/ {k_BT}}}~~,}
where $H$ is the hamiltonian, the trace is
over all states of the system, $k_B$ is Boltzmann's constant
and $T$ is the temperature. More explicitly this may be written as
\eqn\Zsum{{Z}=e^{-E_{GS}/k_B T} \sum_i e^{-(E_i-E_{GS})/k_BT}}
where the sum is over all the eigenvalues $E_i$ of $H$ (with their
multiplicities)
and we have
explicitly factored out the contribution of the ground state
energy $E_{GS}$.

For a macroscopic system we are usually more interested in the free
energy per site $f$ and the specific heat $C$, in the thermodynamic limit
which is defined as
\eqn\TM{{\rm fixed}\quad T>0\quad {\rm \quad and\quad \quad}
 M \rightarrow \infty~.}
The free energy and the specific heat are then
\eqn\sheat{
f=-k_BT\lim_{M \rightarrow \infty}{1\over M}\ln Z~~,~~~~~~~~~
 C~=~-T~{\partial^2 f\over{\partial T^2}}~~.}

Starting with the work of Bethe~\rbethe~in 1931 and
Onsager~\ronsager~in 1944 it has been discovered that there is a
very large number of one-dimensional quantum spin chains (and
two-dimensional classical statistical mechanical systems) whose energy
spectrum and partition function may be exactly studied by what are
essentially algebraic means, starting with the existence of a
family of   commuting   transfer matrices~\rbax
\eqn\comtran{[T(u), T(u')]=0~.}
This commutation relation generalizes the concept of integrability of
classical mechanics. The search for solutions to this equation leads
to the famous star-triangle~\ronsager\rwannier~or
Yang-Baxter equation~\rbax \ryang~and
is beyond the scope of this note. We remark, however, that these
systems can be massive as well as massless, and have profound
connections to the theory of non-linear differential equations~\rwmtb.

\medskip
The next physical ingredient we need is the
approach to the study of
conformal field theories
introduced in 1984 by Belavin, Polyakov, and Zamolodchikov~\rbpz.
The original presentation
is directly relevant for
two-dimensional statistical mechanics.
However, for our present
purpose it is more convenient to formulate the theory in terms of
one-dimensional quantum spin chains. In this formulation, conformal field
theory deals with massless systems whose excitations are characterized
by~\lin~where the $v_\alpha$ are the same for all $\alpha$,
{\it i.e.}~$v_\alpha=v$.
However, instead of the thermodynamic limit~\TM~we study the
limit
\eqn\conlim{M\rightarrow \infty,\quad T\rightarrow 0\quad {\rm
with}\quad MT\quad {\rm fixed},}
which we will refer to as the conformal limit.
Defining the scaled partition function
\eqn\Zhat{ \widehat{Z} \equiv \lim ~e^{Me_0/k_B T} Z~}
in the conformal limit, where $e_0\equiv \lim_{M\to\infty}{1\o M}E_{GS}$,
$\widehat{Z}$ becomes a function
of the    dimensionless   variable
\eqn\qdef{q=e^{-{2 \pi v\over Mk_BT}}~~.}

It is here that the first synthesis with mathematics takes place
because it is found~\rcardy~that
the partition function ${\widehat Z}$ \Zhat\
is expressed in
terms of characters of representations of the Virasoro algebra
(or possibly some extension of it)
 as
\eqn\zfac{{\widehat Z}~=~\sum_{k,l}N_{kl}~\chi_k(q)~\chi_l({\bar q})~~,}
where the $N_{kl}$ are non-negative integers.
In the two-dimensional statistical system ${\bar q}$ is the
complex conjugate of $q$. In the quantum
spin chain context $q=\bar{q}$
is of course real, given by \qdef,
and the factorization corresponds to a decomposition into
contributions coming from the right-moving and left-moving excitations
 separately. This factorization is sometimes called  chiral
decomposition, and the algebras of which $\chi_k$ are characters are
referred to as chiral algebras. In the interpretation as a
two-dimensional statistical system the modular transformation $S$~\gen\
corresponds to interchange of the vertical and horizontal axes. This
interchange should leave the partition function invariant
(when the boundary conditions in both directions are the same),
and this
invariance follows from the modular transformation properties of the
$\chi_k$ if the $N_{kl}$ are suitably chosen~\rcardy \rciz.

If there is only one length scale in the problem, the low-temperature
specific heat computed from~\sheat~should agree with the
specific heat computed from ${\widehat Z}$ of~\zfac~in the limit
$q\rightarrow 1^-$. Generically, if we set $q=e^{2\pi i\tau}$ and
${\tilde q}=e^{-2 \pi i/\tau}$, we have
\eqn\qtoone{\chi_k~\sim ~A_k~{\tilde q}^{{\tilde c}/24}~=~A_ke^{-(2 \pi
)^2{\tilde c}/\ln q}~~~~~~~~{\rm as}~~~q\to 1^-,}
where $\tilde{c}$ is independent of $k$ and the $A_k$ are positive constants
independent of $q$. Using \qtoone\ in \zfac\ one concludes
that the low-temperature behavior of the specific heat
is
\eqn\clowtemp{C~\sim~{\pi k_B {\tilde c}\over 3 v}T~~.}
The quantity ${\tilde c}$ is known as the effective central charge, and
since the $\chi_k$ form a representation of the modular group~\relations\
(with $S$, in particular, transforming $q$ to $\tilde{q}$)
we find that
\eqn\ceff{{\tilde c}=c-24\min_{k} \Delta_k~~.}

\medskip
The final piece of physical information we need is the concept of
Bose-Fermi correspondence in 1+1 dimensions. In
three space dimensions the
concepts of bosons and fermions, whether defined through their spectra
as we did above or in terms of commutation and anti-commutations relations
(which are equivalent definitions due to the spin-statistics connection),
are quite distinct. However, in 1+1 dimensions
they are related. The earliest recognition of such a phenomenon was in
the 1929 paper of Jordan and Wigner~\rjw~and the transformation
they found plays
a key role in the 1949 solution of the Ising model by Kaufman~\rkauf.
In quantum field theory the most familiar example of the phenomenon is
the relation between the massive Thirring model and the sine-Gordon
model~\rcol\rman\rmel.
(Mandelstam's operator~\rman~can be
thought of as implementing in the continuum
 the Jordan-Wigner transformation that relates~\rluther \rluscher~the
spin chains underlying the two field theories.)
This Bose-Fermi correspondence has been extensively studied in the
more general
context of current algebras (see~\rfrishman~for a recent review), and
in the massless case contact is made with affine Lie algebras~\rfrenkel\rwit.

Our goal
here is to indicate that this Bose-Fermi correspondence
is of universal occurrence and that all
conformal field theory
characters
have two   types of sum   representations,
generalizing the right-hand (bosonic) sums of the Rogers-Schur-Ramanujan
identities \szero-\sone\ and their left-hand (fermionic) sums.
The remainder of this
note will concern our recent discoveries of the fermionic counterparts
for large classes of models for which only the bosonic forms have been
known previously.

\bigskip
\newsec{Fermionic sums for conformal field theory characters}

The presentation of the previous two sections strongly suggests that
for solvable one-dimensional quantum spin chains
derived from two-dimensional statistical mechanical
models characterized by commuting
transfer matrices~\comtran, it should be possible to derive the
characters of the related
chiral
algebra by directly
computing the energy levels and the partition function ${\widehat Z}$
\Zhat,
and then putting ${\widehat Z}$ in the form
{}~\zfac. Recently a great deal of progress has been made in this
program for the critical 3-state Potts model~\rkm \rdkmm.
A prominent feature of these methods, which utilize parametrization
of the energy levels in terms of solutions to a set of Bethe equations,
is that they always lead to spectra with the
fermi exclusion rule~\fermi\ and never to bosonic spectra.
Consequently, one obtains fermionic sum representations for the
characters of the chiral algebra of the conformal field theory
which describes the continuum limit of the spin chain.

 The results of the 3-state Potts model computations
strongly suggest further generalizations which were presented
in~\rkkmmone-\rdkkmm. The
most general of these results is that all characters can be written in
the form
\eqn\chiralpart{\widehat{\chi}=\sum_i e^{-\widehat{E}_i/k_B T}~~,}
where
\eqn\enchiral{\widehat{E}_i=
 \sum_{\alpha=1}^n~\sum_{j=1}^{m_{\alpha}}vP_j^{\alpha}~~,}
$n$ is the number of types of quasi-particles, and the $m_{\alpha}$
specify the number of quasi-particles of type $\alpha$ and will in
general be subject to certain restrictions (such as being even or odd).
In addition
\eqn\Pa{P^{\alpha}_{j} \in \Bigl\{ P^{\alpha}_{\rm min}({\bf m}),
  ~P^{\alpha}_{\rm min}({\bf m})+{2\pi \o M},
  ~P^{\alpha}_{\rm min}({\bf m})+{4\pi \o M},
  ~\ldots,
  ~P^{\alpha}_{\rm max}({\bf m}) \Bigr\}~,}
with the further requirement that
the fermi exclusion rule~\fermi\  holds, namely
\eqn\fermirule{P_j^{\alpha}\neq P_k^{\alpha}~~~~\quad{\rm for}\quad j\neq k
  \quad{\rm and~all}\quad \alpha.}
The $P^{\alpha}_{\rm min}({\bf m})$ and
$P^{\alpha}_{\rm max}({\bf m})$ depend linearly on ${\bf
m}=(m_1,m_2,\ldots,m_n)$, with
$P^{\alpha}_{\rm max}({\bf m})$ possibly infinite.

To make the sum \chiralpart\
more transparent,
define $Q_m(N;N')$
to be the number of  additive partitions of $N\geq0$ into $m$ distinct
non-negative integers each less than  or equal to $N'$ (and $Q_m(N)$
to be the number of partitions of $N$ into $m$ distinct non-negative
integers), and recall the identity~\rstanley
\eqn\Qgen{ \sum_{N=0}^\infty Q_m(N;N') ~q^N ~=~ q^{m(m-1)/2}~
  {N'+1 \atopwithdelims[] m}_q~~,}
where the $q$-binomial is defined (for integers $m,n$) by
\eqn\qbin{ {n \atopwithdelims[] m}_q ~=~\cases{ ~~{(q)_n \o (q)_m (q)_{n-m}}
  ~~~~~~~~& if ~~$0\leq m \leq n$ \cr  ~~0 & otherwise.\cr} }
Taking $N'=\infty$ in \Qgen\
 the corresponding expression for $Q_m(N)$ is obtained, namely
 \eqn\sumq{\sum_{N=0}^{\infty}Q_m(N)q^N = {q^{m(m-1)/2}\over (q)_m}~~.}
Thus if   $P^{\alpha}_{\rm min}$ and $P^{\alpha}_{\rm
max}$ are parametrized
in terms of the (symmetric) $n\times n$ matrix $B$ and the $n$-component
vectors ${\bf A}$ and ${\bf u}$ as
\eqn\Pmin{P^{\alpha}_{\rm min}({\bf m})
  ~ =~{2\pi \over M}~{1\o 2} \Bigl[ ({\bf m}(B-1))_{\alpha}- A_{\alpha}+1
 \Bigr]}
and
\eqn\Pmax{ P^{\alpha}_{\rm max}({\bf m}) ~=~ -P^{\alpha}_{\rm min}({\bf m})+
                  {2\pi \o M}({{\bf u}\o 2} -{ \bf A})_{\alpha}~~,}
(and we note that if some $u_\alpha$=$\infty$ the corresponding
$P^\alpha_{\rm max}$=$\infty$) we find from \chiralpart-\Qgen, using \qdef,
that
\eqn\Snauq{\widehat{\chi}= S_B{{\bf Q}\atopwithdelims[]{\bf A}}({\bf u}|q)
{}~\equiv~
 \sum_{{\bf m}\in \ZZ^n \atop {\rm restrictions~ {\bf Q}}}
  q^{{1\o 2}{\bf m} B {\bf m}^t -{1\o 2}{\bf A}\cdot{\bf m}}
  ~\prod_{\alpha=1}^n ~
  { ({\bf m}(1-B)+{{\bf u}\over 2})_\alpha \atopwithdelims[] m_\alpha}_q~~.}
In the special case where all $u_{\alpha}=\infty$ (and hence~\sumq~is
used exclusively in place of~\Qgen) we find that~\Snauq~reduces to
\eqn\SofX{  S_B{{\bf Q}\atopwithdelims[]{\bf A}}(q)
 ~\equiv~ \sum_{m_1,\ldots ,m_n=0
\atop{\rm restrictions~{\bf Q}}}^{\infty} ~
    {q^{{1\over 2}{\bf m} B {\bf m}^t-{1\over 2}{\bf A}\cdot{\bf m} }
 \o (q)_{m_1} \ldots (q)_{m_n} }~~.}

As the simplest example consider the case of $n$=1,
$P_{\rm min}=\pi/M$ or 0, and $P_{\max}$=$\infty,$ which
describes what is called a free chiral fermion
(with anti-periodic or periodic boundary conditions, respectively).
{}From~\Pmin~this
corresponds to $B$=1 and ${\bf A}$=0 or 1 (with no restrictions ${\bf Q}$),
and the corresponding characters \SofX\ are
\eqn\chif{\widehat{\chi}_A^F=
  \sum_{m=0}^{\infty}{ q^{m^2/2}\over (q)_m}~~~,~~~~~~~~~~~
\widehat{\chi}_P^F=
  \sum_{m=0}^{\infty}{ q^{m(m-1)/2}\over (q)_m}~~~.}
These free chiral fermion characters are to be contrasted with the
character of a single chiral boson, computed from \chiralpart-\Pa\
with $n$=1, $P_{\rm min}=2\pi/M$ and $P_{\rm max}$=$\infty$ but
without any exclusion rule on the $P_j$, leading to
\eqn\chib{\widehat{\chi}^B=
 \sum_{N=0}^\infty P(N)q^N = \prod_{n=1}^\infty {1\o 1-q^n}
 ={1\over (q)_{\infty}}~~,}
where $P(N)$ is the number of additive partitions of $N$ into
an arbitrary number of (not necessarily distinct) positive integers.

Upon comparison of~\chib~and~\chif\ with~\szero-\sone~we see
that it is natural to refer to the left-hand sums in~\szero-\sone\
as fermionic and the right-hand sides as bosonic.
To complete the generalization of \szero-\sone\ for the case of a single
free chiral fermion, to exhibit the Bose-Fermi correspondence, and to show
the relation with the
conformal field theory of the Ising model which is the Virasoro
minimal model~\rbpz~${\cal M}(3,4)$,
we note from (83)-(86) of~\rslater~and from~\rrc\rchri~that
\eqn\chioneone{\sum_{m=0\atop {m~{\rm even}}}^{\infty}{q^{{m^2}/2}\over
(q)_m}=
\prod_{n>0\atop n\equiv 2,3,4,5,11,12,13,14({\rm mod}~16)}{1\o 1-q^n}
=q^{1/48}\chi_{1,1}^{(3,4)}}
\eqn\chionethree{\sum_{m=1\atop {m~{\rm odd}}}^{\infty}{q^{{m^2}/2}\over
(q)_m}=
q^{1/2}\prod_{n>0\atop n\equiv 1,4,6,7,9,10,12,15({\rm mod}~16)}{1\o 1-q^n}
=q^{1/48}\chi_{1,3}^{(3,4)}}
\eqn\chionetwo{\sum_{m=0\atop{m~{\rm even}}}^{\infty}
{q^{m(m-1)/2}\over(q)_m}= \sum_{m=1\atop{m~{\rm
odd}}}^{\infty} {q^{m(m-1)/2}\over (q)_m}
=\prod_{n=1}^\infty {1\o 1-q^{2n-1}}
=q^{-1/24}\chi_{1,2}^{(3,4)}}
and thus
\eqn\chifall{\eqalign{
 \widehat{\chi}_A^F &=
  \sum_{m=0}^{\infty}{ q^{m^2/2}\over (q)_m}
 = \prod_{n=1}^\infty (1+q^{n-1/2}) =
 q^{1/48}(\chi_{1,1}^{(3,4)}+\chi_{1,3}^{(3,4)}) \cr
 \widehat{\chi}_P^F &=
  \sum_{m=0}^{\infty}{ q^{m(m-1)/2}\over (q)_m}
 = 2\prod_{n=1}^\infty (1+q^{n}) =
 2 q^{-1/24}\chi_{1,2}^{(3,4)}~~,\cr}}
where
bosonic sum representations for the Virasoro characters on the right-hand
sides are given in \roc.

\medskip

We begin our presentation of fermionic sum representations for characters
with an example where the Bose-Fermi
correspondence (at the level of character formulas)
is particularly
easy to prove: the $SU(2)$  Wess-Zumino-Witten
model~\rwit \rwz. The symmetry algebra of this
conformal field theory
is shown in~\rwit~to be
the affine $su(2)$ Kac-Moody algebra denoted by $(A_1^{(1)})_k$ where
$k=1,2,\ldots$ is the level. At level one
this
theory was argued~\raffleck~to describe the conformal limit of the system
originally studied by Bethe~\rbethe, the spin
${1\over 2}$ Heisenberg anti-ferromagnetic chain
\eqn\heisenberg{H_{{\rm XXX}}=\sum_{j=1}^{M}(\sigma _j^x \sigma _{j+1}^x
+\sigma _j^{y}\sigma_{j+1}^y +\sigma_j^z \sigma_{j+1}^z)}
(where ${\sigma^i}$ are the Pauli spin matrices,
$M$ is even, and periodic boundary conditions $\sigma_{M+1}^i=\sigma_1^i$
are imposed).

One form of the two characters of the $SU(2)$ level 1 theory
was given in~\sucharacters. Comparing this form of the
character with the character
of the free chiral boson~\chib\
we see that~\sucharacters~is interpreted
in terms of
a free chiral boson with an internal quantum number $Q$ (called charge)
that adds an extra term
$2 \pi v Q^2\over M$
to the total energy \enchiral.
The character $q^{1/24}\chi_l$ with $l$=0~(1)
is obtained by summing over all charge sectors with $Q$
an integer (half-odd-integer).
We call this the bosonic form of the
characters.
Product formulas for the characters are readily obtained due
to the fact that \sucharacters\ are two Jacobi theta functions
(divided by $(q)_\infty$), namely
\eqn\suprod{ q^{1/24}\chi_l(q)~ =~ (1+l)~ q^{l/2}
 \prod_{n=1}^\infty (1+q^n)(1+q^{2n+l-1})^2~~~~~~~~~(l=0,1).}

However, it was
shown
by Faddeev and Takhtajan~\rft~that the spectrum
of the spin chain~\heisenberg\
can be constructed from two
fermionic excitations (forming an $SU(2)$ doublet),
and thus a representation of the character in
the form of~\chiralpart\ with $n$=2 should be possible.
Indeed,
we find that the two characters \sucharacters\ have the
representation
\eqn\chixxx{q^{1/24}\chi_l(q)=
\sum_{{m_1,m_2=0\atop
m_1 - m_2
\equiv l({\rm mod}~2)}}^{\infty}
{q^{({m_1+m_2\over 2})^2}\over (q)_{m_1}(q)_{m_2}}~~~.}
When compared with \SofX\ this gives
$B={1\o 2}\pmatrix{1 & 1\cr 1& 1\cr}$ and ${\bf A}$=0,
and thus from~\Pmin\ we see that the minimum momenta
are
\eqn\xxxmin{P^1_{\rm min}({\bf m})={\pi\over M}(1-{m_1-m_2\o 2})~,~~\quad
P^2_{\rm min}({\bf m})={\pi\over M}(1-{m_2-m_1\o 2})~.}

To prove the equality of~\sucharacters~and~\chixxx~we first
recall a relation due to Cauchy (eq.~(2.2.8)
of~\randrewstwo), called the $q$-analogue of Kummer's theorem:
\eqn\qkummer{\sum_{n=0}^{\infty}{q^{n^2-n}z^n\over (q)_n \prod_{j=1}^n
(1-zq^{j-1})} =\prod_{m=0}^{\infty}(1-zq^m)^{-1}~~;}
setting
$z=q^{k+1}$
and dividing by $(q)_k$ we obtain
\eqn\magic{\sum_{n=0}^{\infty}{q^{n^2+nk}\over (q)_n
(q)_{n+k}}={1\over (q)_{\infty}}~~~~~~~~~~(k=0,1,2,\ldots).}
We then write~\chixxx~as
\eqn\start{q^{1/24}\chi_l(q)=
\delta_{l,0}\sum_{m=0}^{\infty}{q^{m^2}\over (q)_m^2}~+~
2\sum_{0\leq m_1<m_2 \atop {m_1-m_2\equiv l({\rm mod}~2)}}^{\infty}
{q^{({m_1+m_2\over 2})^2}\over (q)_{m_1}(q)_{m_2}}~~,}
and set $m_2=m_1+2n-l$ to obtain
\eqn\next{q^{1/24}\chi_l(q)=
\delta_{l,0}\sum_{m=0}^{\infty}{q^{m^2}\over (q)_m^2}+
2 \sum_{n=1}^{\infty}~\sum_{m_1=0}^{\infty}{q^{m_1^2+m_1(2n-l)+{1\over
4}(2n-l)^2}\over (q)_{m_1} (q)_{m_1+2n-l}}~~.}
Then, using~\magic~to reduce the sums over $m$ and $m_1$ we obtain
\eqn\result{q^{1/24}\chi_l(q)={1\over
(q)_{\infty}}(\delta_{l,0}+2\sum_{n=1}^{\infty}
q^{(n-{l\over 2})^2})={1\over(q)_{\infty}}\sum_{n=-\infty}^{\infty}
q^{(n+{l\over 2})^2}}
as desired.

In fact, the
equality of \sucharacters\ and \chixxx~follows from a more
general identity.
For $p$ and $p'$
relatively prime
positive integers, $p\geq p'$, $Q=0,1,\ldots,p-1$ and
$Q'\in {\bf Z}_{2p'}$,
define
\eqn\gdef{G_{Q,Q'}^{(p,p')}(z,q)=
\sum_{m_1,m_2=0\atop {m_1-m_2\equiv Q'({\rm mod}~2p')}}^\infty
{z^{p(m_1-m_2)+Q\over 2pp'}q^{pp'({p(m_1-m_2)+Q\over
2pp'})^2+m_1 m_2}\over (q)_{m_1}(q)_{m_2}}~~.}
Let us also define
\eqn\fdef{f_{a,b}(z,q)={1\over (q)_{\infty}}\sum_{j=-\infty}^{\infty}
z^{j+{b\over 2a}}q^{a(j+{b\over 2a})^2}~~,}
which satisfy the periodicity properties
\eqn\perfab{ f_{a,b}(z,q)=f_{a,b+2a}(z,q)=f_{a,-b}(z,q)~.}
Then exactly the same proof as given above shows that
\eqn\gequalsf{
 G_{Q,Q'}^{(p,p')}(z,q)=f_{pp',pQ'+Q}(z,q).}
The equality of~\sucharacters~and~\chixxx~is just the case
$p=p'=z=1$ of \gequalsf.

Now recall~\rkirits~that $q^{-1/24}f_{pp',pQ'+Q}(1,q)$ form the complete
set of characters of the gaussian $c$=1 model
with compactification radius $r$=$\sqrt{{p\o 2p'}}$ (in the conventions
of~\rgin ). This model is the conformal field theory of
 the XXZ spin chain with the hamiltonian~\ryy
\eqn\xxz{H_{\rm XXZ}=\sum_{j=1}^{M}(\sigma_j^x\sigma_{j+1}^x +\sigma_j^y
\sigma^y_{j+1} +\cos \mu ~\sigma_j^z\sigma_{j+1}^z)}
with
\eqn\radius{r={1\over \sqrt{2(1-{\mu\over \pi})}}~~~~~~~~~~
 (0\leq \mu < \pi).}
The fermi single-particle energies of this model are~\rjkm~(for
$0\leq \mu \leq {\pi \o 2}$)
\eqn\xxzenergy{e(P)={2 \pi \sin \mu\over \mu}\sin P\quad
{}~~~~~~~(0\leq P\leq\pi),}
and so the speed of sound is
$v={2\pi \sin \mu\over \mu}$.
The conformal field theory prediction for the
scaled partition function \Zhat\ of \xxz\ (with $M$ even,
$\sigma_{M+1}^i=\sigma_1^i$,
and $r$=$\sqrt{{p\o 2p'}}$)
is
\eqn\Zxxz{ \eqalign{ \widehat{Z}_{\rm XXZ} &=
 {(q\bar{q})^{-1/24}\o (q)_\infty (\bar{q})_\infty }
{}~\sum_{m,n=-\infty}^\infty q^{{1\o 2}({m\o 2r}+nr)^2}
  \bar{q}^{{1\o 2}({m\o 2r}-nr)^2} \cr
&= (q\bar{q})^{-1/24} \sum_{Q=0}^{p-1}~
 \sum_{Q'=0}^{2p'-1} f_{pp',pQ'+p'Q}(1,q)~f_{pp',pQ'-p'Q}(1,\bar{q})~~.\cr} }

It is suggestive to interpret \gequalsf\ as expressing the Bose-Fermi
correspondence of the gaussian and Thirring models, which are the massless
limits of the sine-Gordon and massive Thirring field theories~\rcol\rmel.
To support this interpretation of the two types of fermionic excitations
in \gdef\ (carrying opposite charge, which is ``measured'' by the
``fugacity'' variable $z$) as the fermion and anti-fermion of the
Thirring model, consider the characters at the Thirring
decoupling point $r$=1, {\it i.e.}~$(p,p')$=(2,1),
which corresponds to the XX point $\mu=\pi/2$ of~\xxz. At this point
we can rewrite
\eqn\decoup{ G_{Q,Q'}^{(2,1)}(1,q)=q^{Q^2/8}
 \sum_{m_1=0}^\infty {q^{m_1(m_1+Q)/2}\o (q)_{m_1}}
 ~\sum_{m_2=0 \atop m_2\equiv m_1+Q'({\rm mod}~2)}^\infty
 {q^{m_2(m_2   -   Q)/2}\o (q)_{m_2}}~~, }
and using the easily obtained identity
\eqn\simid{ \sum_{m=0}^\infty {q^{m(m+1)/2}\o (q)_m} =
  {1\o 2}~\sum_{m=0}^\infty {q^{m(m-1)/2}\o (q)_m} }
we see that the four characters
$q^{-1/24}G_{Q,Q'}^{(2,1)}(1,q)$
 ~($Q,Q'=0,1$) are simple quadratic combinations
of the Ising characters \chioneone-\chionetwo, namely
\eqn\GasI{ \eqalign{
  q^{-1/24}G_{0,0}^{(2,1)}(1,q) = (\chi_{1,1}^{(3,4)})^2 &+
      (\chi_{1,3}^{(3,4)})^2~,~~~~
  q^{-1/24}G_{0,1}^{(2,1)}(1,q) = 2\chi_{1,1}^{(3,4)}
      \chi_{1,3}^{(3,4)} \cr
  q^{-1/24}G_{1,0}^{(2,1)}(1,q) &=
  q^{-1/24}G_{1,1}^{(2,1)}(1,q) = (\chi_{1,2}^{(3,4)})^2~~.\cr}}

{}For points other than $(p,p')$=(2,1) the two fermionic quasi-particles
in \gdef\ do not decouple, which is our interpretation of  the fact
that when bringing $G_{Q,Q'}^{(p,p')}(1,q)$ to the form \SofX\
the matrix $B$ is not diagonal.  Explicitly,
\eqn\bmatrix{B=\pmatrix{{p\over2p'}&1-{p\over 2p'}\cr1-{p\over
2p'}&{p\over 2p'}\cr}~.}
 We note also that the appropriate linear shift in \SofX\
is ${\bf A}=(-Q/p',Q/p')$, and hence the momentum restrictions \Pmin\
in this case are
\eqn\gauspone{ \eqalign{
 P^1_{\rm min}({\bf m})&={\pi\over M}\Bigl[(m_2-m_1)(1-{p\over
2p'})+ {Q\o p'}+1 \Bigr] \cr
 P^2_{\rm min}({\bf m})&=-{\pi\over M}\Bigl[(m_2-m_1)(1-{p\over
 2p'})+{Q\o p'}-1 \Bigr] ~.\cr} }

\medskip
We now are finally in a position to summarize the status of fermionic
representations of conformal field theory characters. All the needed
notation has been introduced above and we may now proceed in a
summary fashion. The results have been originally presented
in~\rkm-\rdkkmm \rter-\rkns. In particular we follow the presentation
of~\rdkkmm. As specified in the original papers some of the
involved $q$-series identities
are
proven and others are conjectures verified to
high orders in $q$.

\subsec{${(G_r^{(1)})_1\times(G_r^{(1)})_1\over (G_r^{(1)})_2}$~ where $G_r$ is
a simply-laced Lie algebra of rank $r$.}

Characters of these coset conformal field theories
are of the form~\SofX\
with
$n=r$
and $B=2C_{G_r}^{-1}$, namely twice the inverse Cartan matrix of $G_r$.
Denoting $S_B{{\bf Q}\atopwithdelims[] {\bf 0}}(q)$ of \SofX\ by
$S_{G_r}^Q(q)$ in this subsection,
the results in the various cases are as follows:

\medskip \no
${\bf G_r = A_n}$: ~This is the original case of Lepowsky and
Primc~\rleppri :
the sums \SofX\ with $B=2C_{A_{N-1}}^{-1}$
are \lpsum-\rest, which provide fermionic sum representations for all
the characters of the corresponding $\ZZ_{n+1}$-parafermionic
conformal field theory~\rZF.
We merely note here
that the linear shift term ${\bf A}_l\cdot{\bf m}$ of \lin\ can be obtained
from
${\bf A}$=0 by replacing
$m_l$ by $m_l+{1\o 2}$
in the quadratic form in \SofX.

\medskip \no
${\bf G_r= D_n~(n\geq 3)}$: The corresponding conformal field theories are
the
points
$r=\sqrt{{n\o 2}}$
on the $c$=1 gaussian line.
Hence, as discussed earlier in this section, the  characters
are $q^{-1/24}f_{n,j}(1,q)$ of \fdef, with $j=0,1,\ldots,n$, for which
a fermionic sum representation in terms of two quasi-particles was
given in \gdef. Now \SofX\ with $B=2C_{D_n}^{-1}$, in the basis where
\eqn\qfDn{ \eqalign{ {\bf m} C_{D_n}^{-1} {\bf m}^t ~=~
  & \sum_{\alpha=1}^{n-2} \alpha m_\alpha^2 ~+~ {n\o 4}(m_{n-1}^2 ~+~ m_n^2)
  ~+~ 2\sum_{1\leq \alpha < \beta \leq n-2} \alpha m_\alpha m_\beta \cr
  &+ ~ \sum_{\alpha=1}^{n-2} \alpha m_\alpha(m_{n-1}+m_n)
  ~+~ {n-2\o 2} m_{n-1} m_n~~,\cr} }
provides a representation for the same characters in terms of $n$
quasi-particles (the degenerate $n$=2 case coinciding with \gdef\
with $(p,p')$=(2,1),
namely \decoup). In particular,
\eqn\SQofDn{ S^Q_{D_n}(q) ~=~ f_{n,nQ}(1,q) }
with $Q=0,1$
indicating  restriction of the
 summation in \SofX\
to
$m_{n-1}+m_n \equiv Q~({\rm mod}~2)$.
Note that due to the coincidence $D_3=A_3$ the expressions \lpsum\ and
\SQofDn\ are related when $n=3$ by (cf.~\rkacwak\rkm )
{}~$S^0_{D_3}=S^0_{A_3}+S^2_{A_3}$~ and ~$S^1_{D_3}=2S^1_{A_3}$.

\medskip \no
${\bf G_r= E_6}$: The conformal field theory is the Virasoro
minimal model~\rbpz~${\cal M}$(6,7) of central charge $c={6\o 7}$
with the $D$-series~\rciz~partition function.
With a suitable labeling of roots we have
\eqn\carEviinv{ C_{E_6}^{-1} ~=~ \pmatrix{4/3 & 2/3& 1& 4/3& 5/3& 2\cr
                                         2/3 & 4/3& 1& 5/3& 4/3& 2\cr
                                         1&  1& 2& 2& 2& 3\cr
                                         4/3& 5/3& 2& 10/3& 8/3& 4\cr
                                         5/3& 4/3& 2& 8/3& 10/3& 4\cr
                                         2& 2& 3& 4& 4& 6\cr} ~,}
and we find (cf.~\roc)
\eqn\SQofEvi{ S^0_{E_6}(q) ~=~ q^{c/24}
  ~[\chi^{(6,7)}_{1,1}(q)+\chi^{(6,7)}_{5,1}(q)] ~~,
   ~~~~S^{\pm 1}_{E_6}(q) ~=~ q^{c/24}~ \chi^{(6,7)}_{3,1}(q) ~~, }
with the restrictions
$m_1-m_2+m_4-m_5 \equiv Q~({\rm mod}~3)$.

\medskip \no
${\bf G_r= E_7}$: ~The conformal field theory is ${\cal M}(4,5)$ of
central charge $c={7\o 10}$. Now
\eqn\carEviiinv{ C_{E_7}^{-1} ~=~ \pmatrix{3/2& 1& 3/2& 2& 2& 5/2& 3 \cr
                                         1& 2& 2& 2& 3& 3& 4 \cr
                                         3/2& 2& 7/2& 3& 4& 9/2& 6 \cr
                                         2& 2& 3& 4& 4& 5& 6 \cr
                                         2& 3& 4& 4& 6& 6& 8 \cr
                                         5/2& 3& 9/2& 5& 6& 15/2& 9 \cr
                                         3& 4& 6& 6& 8& 9& 12 \cr} }
and we find
\eqn\SQofEvii{ S^0_{E_7}(q) ~=~ q^{c/24} ~\chi^{(4,5)}_{1,1}(q) ~~,
   ~~~~S^{1}_{E_7}(q) ~=~ q^{c/24} ~\chi^{(4,5)}_{3,1}(q) ~~, }
when the restrictions  are
$m_1+m_3+m_6 \equiv Q~({\rm mod}~2)$.

\medskip \no
${\bf G_r= E_8}$: ~The coset in this case is equivalent to the Ising
conformal field theory ${\cal M}(3,4)$ of central charge $c={1\o 2}$.
Here
\eqn\carEviiiinv{ C_{E_8}^{-1}~=~\pmatrix{2& 2& 3& 3& 4& 4& 5& 6 \cr
                                          2& 4& 4& 5& 6& 7& 8& 10 \cr
                                          3& 4& 6& 6& 8& 8& 10& 12 \cr
                                          3& 5& 6& 8& 9& 10& 12& 15 \cr
                                          4& 6& 8& 9& 12& 12& 15& 18 \cr
                                          4& 7& 8& 10& 12& 14& 16& 20 \cr
                                          5& 8& 10& 12& 15& 16& 20& 24 \cr
                                          6& 10& 12& 15& 18& 20& 24& 30 \cr} }
and, without any restrictions in the sum \SofX,
\eqn\SofEviii{ S_{E_8}(q) ~=~ q^{c/24} ~\chi^{(3,4)}_{1,1}(q) ~~.}
We further note that
if  $m_1$ in the quadratic form in \SofX\  is replaced
by $m_1-{1\o 2}$, then
one obtains
(up to a power of $q$)
$\widehat{\chi}_{1,1}^{(3,4)}+\widehat{\chi}_{1,2}^{(3,4)}$, and
similarly replacing $m_2$ by $m_2-{1\o 2}$ the combination
$\widehat{\chi}_{1,1}^{(3,3)}+
\widehat{\chi}_{1,2}^{(3,4)}+\widehat{\chi}_{1,3}^{(3,4)}$ is obtained.

\subsec{The cosets of ${(G_r^{(1)})_{n+1}\over U(1)^r}$~.}

This case has been considered in~\rter~and~\rkns~where the identity
characters in the corresponding generalized parafermion conformal
field theory~\rgep~are given by~\SofX\ (with suitable restrictions
on the summation variables) by taking
$B=C_{G_r} \otimes C_{A_n}^{-1}$, which is
explicitly written in a double index notation as
\eqn\Bab{B_{ab}^{\alpha \beta}=(C_{G_r})_{\alpha \beta}(C_{A_n}^{-1})_{ab}\quad
{}~~~~\alpha ,\beta =1,\ldots ,r,\quad a,b,=1,\ldots ,n.}
When $G_r = A_1$, this reduces to the result \lpsum\ of \rleppri.

\subsec{Unitary minimal models ${\cal M}(p,p+1)=
 {(A_1^{(1)})_{p-2}\times(A_1^{(1)})_1\over (A_1^{(1)})_{p-1}}$~.}

For this and subsequent cases we must use the more general form of
eq.~\Snauq.
For  ${\cal M}(p, p+1)$
\eqn\bmat{B~=~{1\o 2}C_{A_{p-2}}~,~~~~~\quad u_1=\infty ~,}
and the ${\bf Q}$-restriction is taken to be $m_a\equiv Q_a$~(mod~2).
Defining
\eqn\QAurs{ \eqalign{ {\bf Q}_{r,s} ~ =~ (s-1)\rhob &+({\bf e}_{r-1}
  +{\bf e}_{r-3}+\ldots)
  + ({\bf e}_{p+1-s} +{\bf e}_{p+3-s}+\ldots)~~\cr}}
where $\rhob=\sum_{a=1}^{p-2} {\bf e}_a$ and
$({\bf e}_a)_b=\delta_{ab}$ for $a=1,\ldots,p-2$ and 0 otherwise,
the conjecture for the
Virasoro characters,
whose bosonic sum representations are given in
\roc, is~\rkkmmtwo
\eqn\main{ \eqalign{ \widehat{\chi}^{(p,p+1)}_{~r,s}(q)~ &=~
  q^{-{1\o 4} (s-r)(s-r-1)}
  S_B{{\bf Q}_{r,s}\atopwithdelims[]{\bf e}_{p-s} }
  ({\bf e}_r +{\bf e}_{p-s}|q)~. \cr}}
Due to
\sym\
another representation must also exist, namely
\eqn\mainalt { \widehat{\chi}^{(p,p+1)}_{~r,s}(q)
 =~q^{-{1\o 4} (s-r)(s-r-1)}~
  S_B{{\bf Q}_{p-r,p+1-s}\atopwithdelims[]{\bf e}_{s-1} }
  ({\bf e}_{p-r}+{\bf e}_{s-1} |q)~~. }

\subsec{Cosets
 ~${(G_r^{(1)})_k\times(G_r^{(1)})_l\over (G_r^{(1)})_{k+l}}$~
 with $G_r$  simply-laced.}

In this case ~$B=C_{G_r}^{-1} \otimes C_{A_{k+l-1}}$, and
 the infinite entries of the vector ${\bf u}$
are $u^{\alp}_l$ for all
$\alp=1,\ldots,r$, in the double index notation used
in subsect.~4.2.

As an example with both $k$ and $l$ greater than 1, consider the case
$G$=$A_1$ with $l$=2, the resulting
series of theories labeled by $k$ being the unitary $N$=1 superconformal
 series whose characters are given in   a   bosonic form
in~\rgko . We find that the character
corresponding to the identity superfield
in these models is obtained by summing over
$m_1\in \ZZ$, $m_a \in 2\ZZ$ for $a=2,\ldots,k+1$.

Another example is the coset
{}~${(E_8^{(1)})_2 \times (E_8^{(1)})_1 \o (E_8^{(1)})_{3}}$~ of
central charge    ~$c= {21\o 22}$, which is
identified as the minimal model ${\cal M}(11,12)$
(with the partition function of the $E_6$-type).
The corresponding sum \Snauq, with ${\bf A}$=0, $u^{\alp}_2$=0
for all $\alp=1,\ldots,8$, and all 16 summations running over all
non-negative integers, gives
{}~$q^{c/24}(\chi_{1,1}^{(11,12)}+\chi_{1,7}^{(11,12)})$, which
is the (extended) identity character of this model.

\subsec{Non-unitary minimal models ${\cal M}(p,p+2)$~ $(p$ odd $)$.}

The character
$\widehat{\chi}^{(p,p+2)}_{(p-1)/2,(p+1)/2}$~  (see \roc) corresponding to
the lowest conformal dimension
{}~$\Delta^{(p,p+2)}_{(p-1)/2,(p+1)/2}=-{3\o 4p(p+2)}$~ in this model
is given by \Snauq\ with ~$B={1\o 2}C_{(p-1)/2}'$ (where
$C_n'$ is defined
at the end of sect.~2),
${\bf A}$=0,
$u_1$=$\infty$ and $u_a$=0 for $a=2,\ldots,{p-1\o 2}$,
and the $m_a$ are summed over all even non-negative integers.

\subsec{Minimal models ${\cal M}(p,kp+1)$.}

For $k$=1 these models are the ones considered in sect.~4.3,
while for $p$=2 they were discussed in sect.~2. Here we
consider the general case.
The character
$\widehat{\chi}^{(p,kp+1)}_{1,k}$ corresponding to the lowest conformal
dimension in the model
is obtained from \Snauq\ with $B$ a $(k+p-3)\times(k+p-3)$
matrix whose nonzero
elements are given by ~$B_{ab}=2(C_{k-1}'^{-1})_{ab}$ ~and
{}~$B_{ka}$=$B_{ak}$=$a$~ for ~$a,b=1,2,\ldots,k-1$, and
{}~$B_{ab}={1\o 2}[(C_{A_{p-2}})_{ab}+(k-1)\delta_{ak}\delta_{bk}]$~ for
{}~$a,b=k,k+1,\ldots,k+p-3$.
Summation is restricted to even non-negative integers
for $m_k,\ldots,m_{k+p-3}$,
the other $m_1,\ldots,m_{k-1}$ running over all non-negative
integers, and
{}~$u_a$=$\infty$ for $a=1,\ldots,k$ and 0 otherwise.

The case $p$=3 is special in that
the fermionic sums are of the form \SofX\ for any $k$.
A slight modification of the matrix $B$ appropriate
for ${\cal M}(3,3k+1)$,
namely just setting $B_{kk}={k\o 2}$ while leaving all other elements
unchanged, gives the
character $\widehat{\chi}_{1,k}^{(3,3k+2)}$~ of ${\cal M}(3,3k+2)$.

\subsec{Unitary $N$=2 superconformal series.}

Expressions
in a
 bosonic form for the characters of these models,
of central charge ~$c={3k \o k+2}$~ where $k$ is a positive integer,
can be found in~\rgep . The identity character,
given by ~$\chi_0^{0(0)}(q)+\chi_0^{0(2)}(q)$~ in the notation
of~\rgep , can be obtained from \Snauq\ if one takes
{}~$B={1\o 2}C_{D_{k+2}}$, ~$u_k$=$\infty$ ~(in the basis used in \qfDn)
and all other $u_a$ set to zero, and $m_{k+1},m_{k+2}$ running
over all non-negative integers while all other $m_a$ summed only over
the even non-negative integers.

\subsec{$\ZZ_N$ parafermions.}

The characters of these models are the branching functions $b^l_m$
given by~\branch, or by the fermionic representation \lpsum.
In sect.~4.3 we found another fermionic representation  for
the case $N$=3 which coincides with the minimal model ${\cal M}(5,6)$
with the $D$-series partition function~\rcardy\rciz.
(The $b^l_m$ in this case are linear combinations of the
$\chi_{r,s}^{(5,6)}$ of \roc.)
This latter form can be generalized to arbitrary $N$. For instance,
$b^0_0$ is obtained from \Snauq\ by setting
{}~$B={1\o 2}C_{D_{N}}$, ~$u_{N}$=$\infty$ ~(in the basis used in \qfDn)
and all other $u_a$ set to zero, and $m_{N-1},m_{N}$ running
over all non-negative integers such that $m_{N-1}+m_{N}$ is even,
while all other $m_a$ are restricted to be even.

\bigskip
\newsec{Discussion}

We have now completed presenting the known results and conjectures for
the fermionic sum representations of
conformal field theory
characters. From these
results a number of questions and speculations arise.

First of all, it is clear that there are many cases where as yet we do
not have any conjectures for the fermionic sums. The most obvious is
the general
Virasoro minimal model ${\cal M}(p,p')$.
Furthermore, for many of the cases
of section 4 not all the characters as yet have conjectured forms. Not
to mention the fact that proofs of the various conjectures remain to
be given.

It would be most useful, however, to turn the program around and to
find the fermionic sum forms directly. For example, it would be highly
desirable to determine mathematically which matrices $B$
in~\Snauq~lead to sums which form a representation of the modular
group~\gen. It should be possible to answer this without
reference to either the bosonic sum or the product representation.
A related question is concerned with the analysis of the leading $q\to 1$
behavior \qtoone\ of the characters, which can be
 obtained~\rkkmmone\rkkmmtwo\rter\rkns\rrisz\rnrt~from their
fermionic sum representations. This analysis gives $\tilde{c}$
of \qtoone\ as a sum of the Rogers dilogarithm function~\rlew~evaluated
at points determined by $B$, and the dilogarithm  sum rules
necessary to reproduce \ceff\ are related~\rnrt\ to deep questions
in different areas of mathematics.

We also want to call the readers attention to the fact repeatedly seen
above that there may be several ``different'' fermionic
representations for the same character. Such a statement is vague
because the concept of ``different'' still remains to be defined.
Nevertheless, as a suggestive specific example we consider the Ising
model characters
\chioneone-\chionetwo.
These three characters were seen in sect.~4.1
to have a representation in terms of one quasi-particle if
the algebra $A_1$ is used, and a representation in terms of eight
quasi-particles if  $E_8$ is used. The representation in
terms of one quasi-particle is
essentially
 the conformal limit of
Kaufman's representation~\rkauf~of the general Ising model in zero
magnetic field in terms of a single free fermion. Similarly, the
representation in terms of eight fermionic quasi-particles  is related
to Zamolodchikov's~\rzam~treatment of the Ising model at $T=T_c$ in a
non-zero magnetic field. It is thus most suggestive to think that the
various ``different'' quasi-particle representations have more than
just a mathematical significance and give insight into the various integrable
massive extensions of conformal field theories.
One further piece of insight is that the
single fermionic quasi-particle
of
the zero-field Ising model has a direct interpretation in the nonlinear
differential  equations that determine the correlation
functions~\rwmtb,
giving rise to a form-factor expansion of the correlation functions.
This work
has been recently extended in the  context of
$N$=2 supersymmetric theories~\rvafa, and it would seem that further
extensions are possible.

We further mention the remarkable fact that the character
formulas \roc\ and \branch\ occur not only in the study of the
spectra of massless systems but also arise in the computation of the order
parameters in off-critical RSOS models by means of corner transfer
matrix techniques~\rbaxtwo-\rfb.
\medskip
However,
some of
these remarks can be considered as tehnical in nature
and
speak
to the mathematical side of the synthesis, but not to the synthesis
itself.
Since our ultimate focus is on the synthesis and not
on computation, we wish to conclude with a few remarks of a more
general nature.

A physicist, as opposed to a mathematician, has an almost inborn
instinct to interpret results in some ``physical'' terms. Inevitably
this process of interpretation involves the setting up of some
catagories
(in the non-mathematical sense),
and even as early as Aristotle it was realized that the
names given to these catagories are not mere labels but carry a great
deal of philosophic content.
This applies also to
 any attempt to
``interpret'' the physical ``meaning'' of the quasi-particle momentum
exclusion rules and the fermionic sum representations presented in this paper.
For example, Haldane~\rhaldane~has attempted to
interpret
momentum exclusion rules similar in spirit to~\xxxmin~in
terms of spinons~\rzou,
``nonorthogonality of localized particle states'', and topological
excitations.
These words are not particularly precise and by their introduction
focus attention on a particular
aspect
of the problem. But  their introduction is
necessitated by the indisputable fact that the excitations which obey
exclusion rules discussed above cannot be
described in the language of conventional second quantization.

In the physical
context of the fractional quantum Hall effect Haldane credits
Laughlin~\rlaughlin~with the realization that second quantization
fails to appropriately
describe the observed phenomena.
However, need to invent new concepts is much more widespread than this
particular area of condensed matter physics. In particle physics it
has been recognized for at least 30 years that the conventional
second-quantized quantum field theory which describes point-like particles
has severe shortcomings.
For example, the discovery of confinement in Quantum
Chromo-Dynamics demonstrates the need to go beyond this concept.
The original motivation for the construction of string theory was to
understand
the strong interactions in what later became known as
the confining phase of QCD
(see~\rgross~for a recent discussion).
This need to go beyond second-quantized point field theory has been
extensively investigated not only at the level of hadrons, but at the
more fundamental level of unifying string interactions with quantized
gravity. It is a most remarkable coincidence that
the mathematics
considered in this paper also occurs in these studies of string theory.
Such a
coincidence cannot be accidental and the fact that mathematicians,
high-energy physicists, condensed matter physicists, and physicists
studying statistical mechanics are all
contemplating the same abstract object is a truly remarkable
demonstration that the whole is much more than the sum of its parts.
The synthesis will be achieved when language can be developed that
incorporates all aspects of the phenomena at the same time.

\bigskip
\bigskip

\noindent
{\bf Acknowledgements}

\smallskip
We wish to thank Prof.~S.T.~Yau for the opportunity of contributing to
these proceedings,
and Dr.~S.~Dasmahapatra,
Dr.~T.R.~Klassen, Prof.~J.~Lepowsky
 and  Prof.~L.A.~Takhtajan
for fruitful discussions. The work of RK and BMM is
partially supported by the National Science Foundation under grant
DMR-9106648, and that of EM under NSF grant 91-08054.

\vfill

\eject
\listrefs

\vfill\eject

\bye
\end